\documentclass{article}

\usepackage{arxiv}

\usepackage[utf8]{inputenc} 
\usepackage[T1]{fontenc}    
\usepackage{hyperref}       
\usepackage{url}            
\usepackage{booktabs}       
\usepackage{amsfonts}       
\usepackage{nicefrac}       
\usepackage{microtype}      
\usepackage{lipsum}
\usepackage{graphicx}
\usepackage{cite}
\usepackage{amsmath,amssymb,amsfonts}
\usepackage{algorithm,algorithmic}
\graphicspath{ {./images/} }
\def\BibTeX{{\rm B\kern-.05em{\sc i\kern-.025em b}\kern-.08em
    T\kern-.1667em\lower.7ex\hbox{E}\kern-.125emX}}

\title{Causality-Driven Reinforcement Learning for Joint Communication and Sensing}

\author{
 Anik Roy \\
  Indian Institute of Technology Kharagpur \\
  Kharagpur 721302, India\\
  \texttt{anikroy@kgpian.iitkgp.ac.in} \\
   \And
 Serene Banerjee \\
  Ericsson Research \\
  Bengaluru, India\\
  \texttt{serene.banerjee@ericsson.com} \\
  \And
  Jishnu Sadasivan \\
  Ericsson Research \\
  Bengaluru, India\\
  \texttt{jishnu.sadasivan@ericsson.com} \\
  \And
   Arnab Sarkar \\
  Indian Institute of Technology Kharagpur \\
  Kharagpur 721302, India\\
  \texttt{arnab@atdc.iitkgp.ac.in} \\
  \And
 Soumyajit Dey \\
  Indian Institute of Technology Kharagpur \\
  Kharagpur 721302, India\\
  \texttt{soumya@cse.iitkgp.ac.in} \\
}

\begin{document}
\maketitle
\begin{abstract}
The next-generation wireless network, 6G and beyond, envisions to integrate communication and sensing to overcome interference, improve spectrum efficiency, and reduce hardware and power consumption. Massive Multiple-Input Multiple Output (mMIMO)-based Joint Communication and Sensing (JCAS) systems realize this integration for 6G applications such as autonomous driving, as it requires accurate environmental sensing and time-critical communication with neighboring vehicles. Reinforcement Learning (RL) is used for mMIMO antenna beamforming in the existing literature. However, the huge search space for actions associated with antenna beamforming causes the learning process for the RL agent to be inefficient due to high beam training overhead. The learning process does not consider the causal relationship between action space and the reward, and gives all actions equal importance. In this work, we explore a causally-aware RL agent which can intervene and discover causal relationships for mMIMO-based JCAS environments, during the training phase. We use a state dependent action dimension selection strategy to realize causal discovery for RL-based JCAS. Evaluation of the causally-aware RL framework in different JCAS scenarios shows the benefit of our proposed framework over baseline methods in terms of the beamforming gain.
\end{abstract}


\section{INTRODUCTION}
\label{secIntro}
The use of a large number of antennas at the transmitter and receiver to enable Massive Multiple-Input Multiple Output (mMIMO) is a key characteristic of future generation Joint Communication and Sensing (JCAS) systems as it significantly compensates for path loss and ensures sufficient signal strength. These systems use analog beamformers where the transceivers employ networks of phase shifters~\cite{Alkhateeb2014} to enable antenna beamforming of beam codebooks. Traditionally, beam codebooks consists of a large number of single-lobe directional beams which are used for data transmission and/or sensing~\cite{Alrabeiah2020}. However, the design of beam codebooks is challenging due to the highly dynamic nature of wireless channels. Furthermore, the beam design criteria for communication and sensing functions are different. For example, a communication beam is designed to focus the transmitter power towards the receiver, while a sensing beam is designed to focus on sensing targets in the environment~\cite{Demirhan2022}. This requires that the beamforming technique adapts to its environment in relation to user distribution and sensing targets, which can potentially be achieved if the mMIMO system incorporates a data-driven artificial intelligence component. Traditionally, Reinforcement Learning (RL)-based techniques have been employed for JCAS-based optimized beamforming as they can account for the complex interactions between the environment and the mMIMO antenna system~\cite{Xu2022, Xu2023}. 

However, traditional RL techniques suffer from high beamforming training overhead to cover all possible directions using a large number of beams in the beam codebook. This high training overhead is mainly due to the difficulty in exploration caused by the large action space and state space of the beamformers in the beamforming codebook. Techniques such as immitation learning~\cite{Sun2018},~\cite{Vinyals2019} cannot be used for the reduction of exploration space in the context of wireless communication due to the highly dynamic nature of the state space.
The dynamic nature of the problem, with its larger state and action spaces, makes it computationally expensive.

We address the gap in the state-of-art as follows; the agent, that learns the antenna beam pattern, needs to discover the causal relationship for the environment to select useful action space during policy training to improve learning efficiency. In this work, we propose the use of the State-Wise Action-Refined Temporal Difference Learning algorithm with IC-INVASE framework (TD3-INVASE) presented in~\cite{Sun2022} to conduct interventions on the action space during training for discovering the causal relationship between the action space and the reward in a mMIMO-based JCAS setup. This model enables the selection of relevant actions based on causal discovery, which will facilitate sample efficient training for extremely large action and state space applications like mMIMO-based JCAS. In addition to capturing the causal relationships via TD3-INVASE, the framework is also capable of capturing causal relationships from expert and domain knowledge.

This is the first work that incorporates causal discovery for efficient state space exploration in mMIMO-based JCAS applications. The State-Wise Action-Refined Temporal Difference Learning framework~\cite{Sun2022} intervenes the learning process and selects the relevant actions based on the causal discovery. This reduces the exploration for extremely large action space such as in case of JCAS applications. In this context, our proposed model is the first work that incorporates causal discovery in mMIMO-based JCAS.

The main contributions of this paper are as follows:
\begin{enumerate}
    \item We propose a novel RL-based framework for beamforming in JCAS environments that learns codebook beam patterns for both communication and sensing based on different environments, user distributions, array geometry and sensing target locations.
    \item In this context, we explore the use of causal RL based neural learning architectures like TD3-INVASE~\cite{Sun2022} that enables causal relationship learning between actions and states, leading to exploration of exremely high dimensional beamforming action space.
    \item Though we demonstrate our approach using TD3-INVASE, the framework allows the capture of causal relationships via expert, domain knowledge, known correlations of the radio propagation environment, or other approaches as well.
    \item For validating our approach, we consider a diverse set of mobility scenarios generated using well known benchmarking tools. Our experiments establish that causality ascription between state and action spaces in the context of JCAS-based beamforming provides for sample efficient training and higher beamforming gain in comparison to baseline RL training policies employed for communication-only beamforming~\cite{Zhang2021}.
\end{enumerate}

\textit{Related Works}: The idea of using RL for beamforming in JCAS applications have been extensively explored in the recent studies~\cite{Pham2022},~\cite{Ahmed2021},~\cite{Zhang2020},~\cite{Zhang2021},~\cite{Xu2022},~\cite{Xu2023},~\cite{Mateos2023},~\cite{Chen2023}. The works~\cite{Pham2022},~\cite{Ahmed2021} investigate the beamforming design problem and proposes value-based reinforcement learning models to optimize the beam pattern. The work reported in~\cite{Ahmed2021} uses the deep Q-learning model to optimize the beamforming vector for achieving the best beamforming gain. However, size of the deep Q-learning model keep increasing as the action dimension increases, thus making it infeasable for mMIMO systems. To address this issue, the authors in~\cite{Zhang2020},~\cite{Zhang2021} proposed the use of Wolpertinger-based architecture~\cite{Arnold2015} for Deep Reinforcement Learning (DRL). Their proposed models can autonomously learn the beamforming codebook to best match the surrounding environment, user distribution, and antenna array geometry and achieve higher beamforming gain than pre-defined classical codebooks. Extending this to JCAS, Lifan \textit{et. al.} in~\cite{Xu2022} proposed a method that uses the Wolpertinger architecture-based framework for beamforming vector learning in order to enhance the performance of automotive radar-based JCAS. Their model can intelligently adjust its phase shifters to steer the beam for tracking a target of interest or enhance communication capacity, while addressing the large dimensionality issue of the action space. Similarly, the authors in~\cite{Xu2023} propose an RL-based framework that jointly optimizes the beam design, power and computational cost for JSAC applications, by incorporating antenna selection in each channel use. Chen \textit{et. al.} in~\cite{Chen2023} proposed a beam-tracking mechanism for JCAS using Deep Neural Network (DNN) to predict the beam direction by learning channel correlations with the help of aggregated sensing information from multiple radars. Furthermore, the authors in~\cite{Mateos2023} propose a model-driven learning architecture for optimizing the beam design for MIMO sensing and multiple-input single-output (MISO) communication. Although, the prior works on beamforming in JCAS have addressed the issue of high action dimensionality in terms of architectural modeling, the traditional exploration methods lead to significant training overhead and sub-optimal beam pattern design, particularly in highly dynamic wireless environments. Furthermore, design of optimum beamforming solutions involve prohibitive computational overheads due to the necessity to exhaustively explore huge state and action spaces.
In this work, we propose an RL technique founded on TD3-INVASE~\cite{Sun2022} for a dynamic mMIMO-based JCAS setup. 

The rest of the paper is organized as follows. Section~\ref{secProbForm} outlines the overall problem formulation. Section~\ref{secCausalBeam} and Section~\ref{secBeamCodebook} presents the proposed framework for RL-based JCAS with causal discovery. Experimental environment setup and results have been presented in Section~\ref{secExpSetup} and Section~\ref{secResults}, respectively. The concluding remarks are given in Section~\ref{secConc}.

\section{BACKGROUND}
\label{secBackgrnd}
\textbf{Reinforcement Learning: } In general, a RL task is defined by a Markov Decision Process (MDP)~\cite{Otterlo2012} denoted by a tuple $\langle \mathcal{S},\;\mathcal{A},\;\mathcal{P},\;r,\;\gamma\rangle$, where $\mathcal{S}$ is the state space, $\mathcal{A}$ is the action space, $\mathcal{P}$ is the transition probability function of the environment, $\mathcal{R}$ is the reward function, and $\gamma \in [0,1)$ is the discount factor. A policy $\pi \in \Pi$ is a probability function that maps states to actions, i.e., $\pi(S) \rightarrow P_r(\mathcal{A})$, where $\Pi$ is the set of all policies. At each time instant $t$, the agent in state $s_t \in \mathcal{S}$, takes an action $a_t \in \mathcal{A}$ given by the policy $\pi(s_t)$, transitions to state $s_{t+1} \sim \mathcal{P}(.|s_t,a_t)$, and receives a reward $r_t = \mathcal{R}(s_t,a_t)$. The objective of the RL algorithm is to learn a policy $\pi^* \in \Pi$ that maximizes the agent's expected cumulative reward $\mathbb{E}_{\tau \sim \pi}[\sum_{t=1}^{\mathcal{\infty}} \gamma^t r_t]$, where $\tau$ is the trajectory obtained from $\pi$ under the influence of the transition dynamics $\mathcal{P}$.

In many environments, it becomes impossible to model the transition probability function $\mathcal{P}$ as all the state variables may not be visible to the agent. This makes design of the MDP infeasible.
Model-free learning allows the agent to make decisions based on experiences from interactions with the environment, treated as a \emph{black box}. The reward function $\mathcal{R}$ acts as a \emph{state-action quality evaluator}, returning a value that reflects the \emph{quality} of the state-action pair.
In Deep Reinforcement Learning (DRL), neural networks are used to represent policy $\pi(s)$ and the \emph{state-action quality evaluator} $Q(s,a)$. One prominent DRL approach is the actor-critic framework~\cite{Sutton2018}. In the actor-critic model, the policy $\pi$, which maps the agent's state $s_t \in \mathcal{S}$ to action $a_t \in \mathcal{A}$, is represented by the actor network with parameter $v$ as $\pi_v$. The critic network with parameter $\phi$ approximates the estimated return for the current state-action pair $(s_t,a_t)$ as $Q_\phi^\pi(s_t,a_t) = \mathbb{E}_{(s_t,a_t) \sim \pi_v,\tau}[\sum_{k=1}^{\mathcal{\infty}} \gamma^k r_{t+k}]$. Each interaction of the agent with the environment is stored in the replay buffer $\mathcal{B}$ as a tuple $\langle s_t,a_t,r_t,s_{t+1} \rangle$. For a set of tuples $\langle s_t,a_t,r_t,s_{t+1} \rangle \sim \mathcal{B}$, the critic is optimized to minimize the Temporal Difference (TD) error~\cite{Watkins1992},
\begin{align}
\nonumber
    \mathcal{L_{TD}} &= \mathbb{E}_{\langle s_t,a_t,r_t,s_{t+1} \rangle \sim \mathcal{B}}[(r_t + \gamma Q_\phi^\pi(s_{t+1},\pi_v(s_{t+1})) \\
    &-Q_\phi^\pi(s_t,a_t))^2]
\end{align}
The reward estimate $Q_\phi^\pi(s_t,a_t)$ from the critic network is used to update the policy (actor) network by gradient descent such that the critic's output is maximized, i.e., $\pi_v(s_t)=\arg\max_{a_t}Q_\phi^\pi(s_t,a_t)$. The gradient of the policy network is given by $\mathbb{E}_{s=s_t,a=\pi_v(s_t)}[\nabla_aQ(s,a)\nabla_v\pi_v(s)]$.

\textbf{INVASE:} This particular learning architecture was proposed in~\cite{Yoon2018} to perform instance-wise feature selection to reduce over-fitting in predictive models by discovering a subset of most relevant features at each sampling instance. A selection function is learned by minimizing a Kullback-Leibler (KL) divergence $\mathcal{D}_{KL}$ between the conditional probability distribution for all features $p(Y|X=x)$ and minimal-selected-features $p(Y|X^{(F(x))=x^{F(x)}})$, which is represented as,
\begin{align}
\label{eqINVASELoss}
\nonumber
    min_F \mathcal{L} &= \mathcal{D}_{KL}(p(Y|X=x)||p(Y|X^{(F(x))=x^{F(x)}})) \\
    &+ \lambda|F(x)|_0
\end{align}

Here, $X$ is the input feature and $Y$ is the predicted output. Function $F:\mathcal{X} \rightarrow \{0,1\}^d$ is a feature selection function, $|F(x)|_0$ denotes the number of selected features, $d$ is the dimension of input features, and $x^{(F(x))} = F(x) \odot x$ denote the element-wise product of $x$ with the selection function $F(x)$. The KL divergence for any random variables $W$ and $V$ with densities $p_W$ and $p_V$ is defined as $\mathcal{D}_{KL}(W||V) = \mathbb{E}[log\frac{p_W(W)}{p_V(V)}]$. An optimal selection function $F$ can minimize Equation~\ref{eqINVASELoss}. INVASE uses the actor-critic framework to optimize the $F$. The selection function $F$ is represented by a neural network $f_\theta(.|x)$, parameterized by $\theta$, and uses a stochastic actor to generate probabilities for each feature. Prediction and baseline networks (used for variance reduction) are trained to return the loss function $\mathcal{L}$, based on which $f_\theta(.|x)$ is optimized using gradient descent $\mathbb{E}_{(x,y) \sim p}[\mathcal{L}\nabla_\theta log f_\theta(.|x)]$.

\section{PROBLEM FORMULATION}
\label{secProbForm}
\begin{figure}[htbt!]
\centering
  \includegraphics[width=0.7\linewidth,clip]{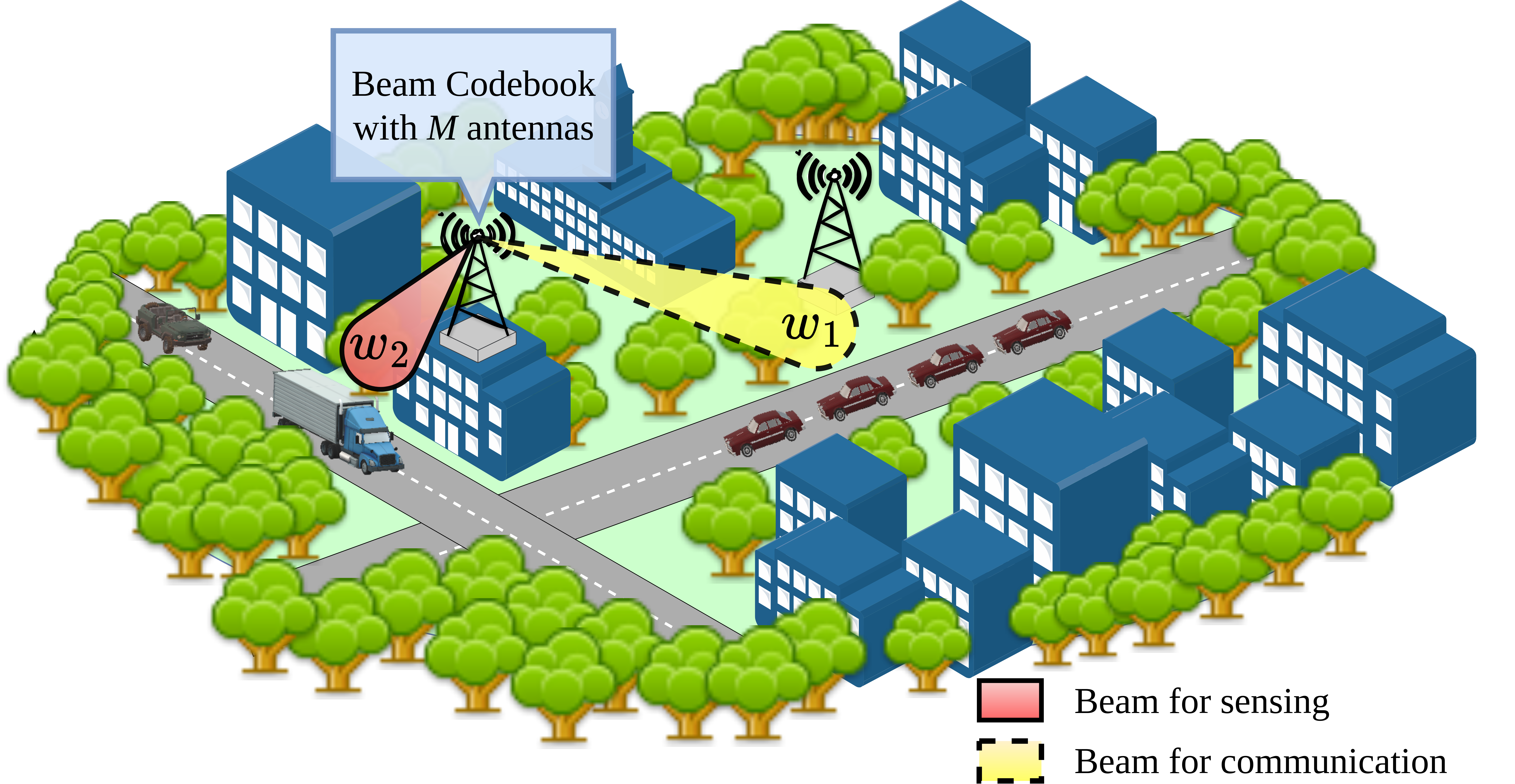}
   \caption{A mMIMO base station with antenna array of $M$ antennas using beam codebook to serve communication users and sense targets in vicinity}
   \label{figScenario}
\end{figure}

We consider a scenario depicted in Fig.~\ref{figScenario}, where a mMIMO base station with $M$ number of antennas is communicating with single antenna users (fleet of vehicle users in Fig.~\ref{figScenario}) and is sensing targets in the environment. The fleet of vehicles are equipped with lidar and radar systems that can only sense Line-of-Sight (LOS) traffic. Therefore, incoming traffic from Non-Line-of-Sight (NLOS) can be potential hazards to these vehicle users. The incoming NLOS traffic are sensing targets for the base station, so that the vehicle users are notified about the potential hazard beforehand. The base station uses analog-only beamformers with a network of $r$-bit quantized phase shifters. Let the beamforming codebook $W$ adopted by the base station consist of $N$ beamforming vectors, where each vector takes the form
\begin{equation}
    \label{eqBfvec}
    w_i = \frac{1}{\sqrt{M}}[e^{j\theta_1}, e^{j\theta_2}, ..., e^{j\theta_M}]^T, for\;i = 1,2, ..., N
\end{equation}

\noindent where each $\theta_m$ phase shifter takes the values from a finite set of $2^r$ possible discrete values drawn uniformly from ($-\pi,\pi$]. Let $N_C$ be the number of communication beams and $N_S$ be the number of sensing beams in the $N$-sized beam codebook. Fig.~\ref{figScenario} depicts dashed beam ($w_1$) to serve communicating users and solid beam ($w_2$) to sense potential hazards. Here, $w_1$ is beamforming vector responsible for communication, while $w_2$ is the vector responsible for sensing.

Let us consider that the communicating users do uplink transmission by sending some signal $x$ to the base station with power $\mathbb{E}[|x|^2]$. Then, the received signal at the base station after combining with the beamforming vector $w_i$ can be expresses as $y_u = w_i^Hh_ux + n$, where $h_u \in \mathbb{C}^{M \times 1}$ is the uplink channel vector between the user (or multiple users with similar channel) $u$ and the base station antennas, and $n \sim \mathcal{N}_\mathbb{C} (0,\sigma_n^2I)$ is the received noise vector at the base station~\cite{Marzetta2016}. Here, $i \in \mathcal{C}$, where $\mathcal{C}$ is the set of communication beam indices in the beam codebook and $|\mathcal{C}|=N_C$. The beamforming gain of the communication beamformer $w_i$ for user $u$ is $g_u=|w_i^H h_u|^2,\forall i \in \mathcal{C}$.

In the context of sensing, the base station antennas’ array response is given as
\begin{equation}
    \label{eqAntArray}
    b_r = [e^{\frac{j2 \pi d_1 sin\theta}{\lambda}},e^{\frac{j2 \pi d_2 sin\theta}{\lambda}},...,e^{\frac{j2 \pi d_M sin\theta}{\lambda}}]^T
\end{equation}

\noindent where $d_1,d_2, ... ,d_m,m=1,2, ... ,M$ denotes the transmit antenna positions. The angle $\theta$ is the Angle of Arrival (AoA) of the echo signal for sensing from the target. The received sensing gain at the base station is given as $g_r=|w_i^H b_r|, \forall i \in \mathcal{S}$. Here, $\mathcal{S}$ is the set of sensing beam indices in the beamforming codebook and $|\mathcal{S}|=N_S$.

Our objective is to learn optimal beamforming vectors $w_{i,opt}, \forall i \in \mathcal{C}\cup\mathcal{S}$ for mMIMO-based JCAS. We formulate a joint optimization problem where the communication beam learning problem can be formulated as,
\begin{align}
    \label{eqCommBeamOpt}
    \nonumber
    w_{i,opt} &= \underset{w_i}{\mathrm{argmax}} \frac{1}{|\mathcal{H}_s|} \sum_{h_u \in \mathcal{H}_s} |w_i^Hh_u|^2, \\
    \nonumber
    &s.t. \text{ } \forall i \in \mathcal{C}, \\
    \textbf{w}_m &= \frac{1}{\sqrt{M}}e^{j\theta_m},\theta_m \in \Theta, \forall m = 1,2,...,M,
\end{align} 

\noindent and the sensing beam learning problem can be formulated as,
\begin{align}
    \label{eqSenseBeamOpt}
    \nonumber
    w_{i,opt} &= \underset{w_i}{\mathrm{argmax}} |w_i^Hb_r|^2, \\
    \nonumber
    &s.t. \text{ } \forall i \in \mathcal{S}, \\
    \textbf{w}_m &= \frac{1}{\sqrt{M}}e^{j\theta_m},\theta_m \in \Theta, \forall m = 1,2,...,M,
\end{align}

\noindent where $w_{i,opt}$ is the optimum beamforming vector for communication beams $\forall i \in \mathcal{C}$ and sensing beams $\forall i \in \mathcal{S}$, $\textbf{w}_m$ is the $m^{th}$ element in the beamforming vector $w_i$, $\Theta$ is the discrete set of possible phase values and $\mathcal{H}_s$ is the channel set that contains single or multiple similar channels.

\section{DRL BEAM CODEBOOK LEARNING FOR JCAS}
\label{secBeamCodebook}
\begin{figure}[htbt!]
\centering
  \includegraphics[width=0.7\linewidth,clip]{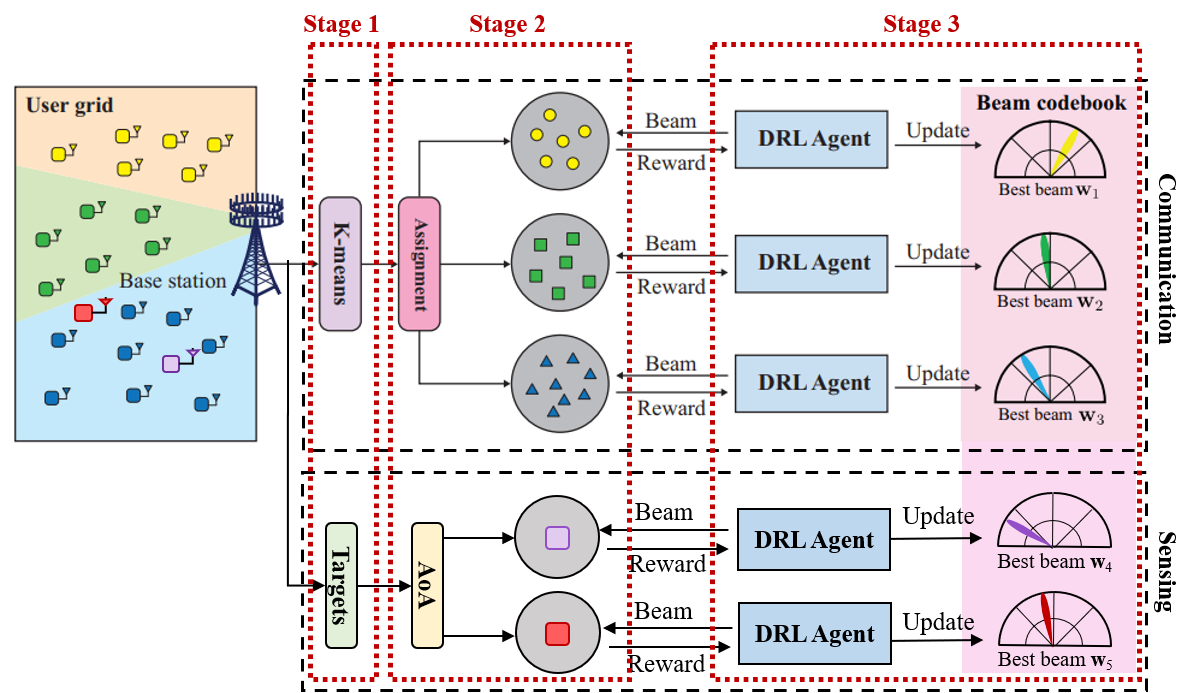}
   \caption{The proposed beam codebook design framework for mMIMO JCAS using DRL. The codebook design framework for communication has been presented in~\cite{Zhang2021}. We extend that model to JCAS in our proposed framework by adding beam design stages for sensing.}
   \label{figSysModel}
\end{figure}

In this section, we describe our multi-network DRL architecture to optimize the JCAS beam learning problem formulated in the previous section with reference to the Fig.~\ref{figSysModel}. The work~\cite{Zhang2021} has presented a three stage beam codebook learning framework for communication. We extend their model to JCAS by incorporating a three stage pipeline for sensing beam learning in our proposed framework. We will first discuss the stages for communication that has been proposed in the existing literature. Then we will discuss our proposed approach for sensing beam design in order to realize JCAS.

\subsection{COMMUNICATION BEAM DESIGN}
\subsubsection{STAGE 1}
The first stage partitions or clusters the users in the environment based on channel similarity. The clustering method relies on a sampled subset $\mathcal{F} = \{f_1,f_2,...,f_C\}$ of the communication beam vectors (Equation~\ref{eqBfvec}) to gather information about the user distribution in the environment in the form of received gain, where $f_s \in \mathbb{C}^M,\;\forall s \in \{1,...,C\}$. Let $K^\prime$ channels contribute to the received gain in the clustering process and they are denoted by $\mathcal{H}_{Cl} = \{h_1,h_2,...,h_{K^\prime}\}$, where $\mathcal{H}_{Cl} \subseteq \mathcal{H}_S$. The received gains are used to construct a matrix $P$ given as,
\begin{equation}
    P = \begin{bmatrix}
        |f_1^Hh_1|^2 & \cdots & |f_1^Hh_{K^\prime}|^2 \\
        \vdots & \ddots & \vdots \\
        |f_C^Hh_1|^2 & \cdots & |f_C^Hh_{K^\prime}|^2
    \end{bmatrix}
\end{equation}

Here, each column represents the received gains of each user cluster in the environment. Applying a clustering algorithm directly on $P$ has been empirically shown to yield overlapping clusters that are hard to separate~\cite{Zhang2021}. Therefore, the authors in~\cite{Zhang2021} propose to construct a feature matrix $U = [u_1,u_2,...,u_{K^\prime}]$ by finding the pair-wise differences of the columns in matrix $P$, which is constructed as follows,
\begin{align}
\nonumber
    u_k = &(\frac{1}{C}\sum_{c=1}^C|f_C^Hh_k|^2)^{-1}
    \begin{bmatrix}
        |f_1^Hh_k|^2 - |f_2^Hh_k|^2 \\
        \vdots  \\
        |f_{C-1}^Hh_k|^2 - |f_C^Hh_k|^2
    \end{bmatrix} \\
    &\forall k \in \{1,2,...,K^\prime\}
\end{align}

Now, the clustering is applied on the columns of the matrix $U$ to form $N_C$ channel (user) clusters using K-means algorithm~\cite{Bishop2006}. The columns $u_k$, where $k=1,2,...,K^\prime$, are C-dimensional vectors given as input to the \emph{distortion measure} objective function $J=\sum_{k=1}^{K^\prime}\sum_{n=1}^{N_C}r_{kn}||u_k-\mu_n||^2$, where $\mu_n,\;\forall n \in \{1,...,N_C\}$ is a C-dimensional vector representing the centre of the $n^{th}$ cluster and $r_{kn} \in \{0,1\}$ indicates whether $u_k$ is assigned to cluster $\mu_n$. The objective function $J$ is minimized in two phases, (a) re-assignment of columns $u_k$ to clusters, and (b) re-computing the cluster centres $\mu_k$, until there is no further change in clustering assignment. The channels in $\mathcal{H}_{Cl}$ are partitioned into $N_C$ disjoint sets as $\mathcal{H}_{Cl}=\mathcal{H}_1 \cup \mathcal{H}_2 \cup ... \cup \mathcal{H}_{N_C}$, where $\mathcal{H}_k \cap \mathcal{H}_l=\Phi, \forall k\neq l$.

\subsubsection{STAGE 2}
The second stage assigns each of $N_C$ channel clusters to $N_C$ different DRL agents, where each DRL learns the communication beam for the assigned channel cluster. Let $\hat{\mathcal{H}}_{Cl}=\hat{\mathcal{H}}_1 \cup \hat{\mathcal{H}}_2 \cup ... \cup \hat{\mathcal{H}}_{N_C}$ be the new cluster assignment based on the k-means clustering algorithm discussed before. Let $\Gamma = \{\hat{w}_1,...,\hat{w}_{N_C}\}$ be the "temporarily best" beamforming vectors of each of the $N_C$ DRL agents assigned to corresponding channel clusters in $\hat{\mathcal{H}}_{Cl}$. The goal is to find the cluster-network assignment that minimizes a cost sum which is a function of the average beamforming gain calculated using the newly assigned channels $\hat{\mathcal{H}}_{Cl}$ and current beamforming vector $\Gamma$. The cost function is calculated as $Z_{nn^\prime}=\frac{1}{\hat{\mathcal{H}}_{n^\prime}}\sum_{h\in \hat{\mathcal{H}}_{n^\prime}}|\hat{w}_n^Hh|^2$,
where $Z_{nn^\prime}$ is the average beamforming gain of the $n^{th}$ temporarily best beamforming vector in $\Gamma$ with the $n^\prime$-th channel cluster in $\hat{\mathcal{H}}_{Cl}$. Using the cost matrix, the clustering assignment is formulated as a linear sum assignment problem,
\begin{align}
    &\min_{X} -\sum_{n=1}^{N_C}\sum_{n^\prime=1}^{N_C}X_{nn^\prime}Z_{nn^\prime} \\
    &s.t.\;X\text{ is a permutation matrix}
\end{align}

This problem is solved using Hungarian algorithm~\cite{Kuhn1955} to obtain $N_C$ tuples which associates each beamforming vector to the channel cluster as $(\hat{w}_n,\hat{\mathcal{H}}_{n^\prime}), n,n^\prime \in \{1,2,...,N_C\}$.

\subsubsection{STAGE 3} 
After channel clustering and their assignment, each channel cluster is used by a DRL agent to train its own beam. The user clustering and assignment are two key stages that are performed at frequent intervals in order to enable adaptibility to dynamic environment. The beam pattern learning at each of the $N_C$ DRL agents proceed according to the procedure which we will discuss in Section~\ref{secCausalBeam}, with an additional fine-tuning step involving perturbation of the beam vector with exploration noise, quantization of the vector, and it's evaluation on the assigned user cluster.

\subsection{SENSING BEAM DESIGN}
\subsubsection{STAGE 1}
The first stage involves identifying the sensing targets. Let there be $\mathcal{T} = \{t_1,t_2,...,t_{N_S}\}$ targets that need to be sensed in the environment and tracked by the sensing beams of the base station. We provide our model the initial location of each of the targets in $\mathcal{T}$ as coordinates $(x_i,y_i),\;\forall i\in \{1,2,...,N_S\}$.

\subsubsection{STAGE 2}
The second stage calculates the AoA $\theta_i, \forall i \in \{1,2,...,N_S\}$ at the antenna steering vector (Equation~\ref{eqAntArray}) located on the base station. The AoA for each target in $\mathcal{T}$ is obtained as
\begin{align}
    \theta_i = \tan^{-1}\Big(\frac{y_i - y_{bs}}{x_i - x_{bs}}\Big),\;\forall i \in \{1,2,...,N_S\}
\end{align}

where $(x_{bs},y_{bs})$ is the coordinate of the base station. We obtain AoAs for all targets $\Theta = \{\theta_1,\theta_2,...,\theta_{N_S}\}$, where the index of $\theta$ is indicative of the target index in $\mathcal{T}$. Each target in $\mathcal{T}$ is assigned with a DRL agent for sensing beam learning.

\subsubsection{STAGE 3}
The third stage learns the beam pattern for each of these targets, where each DRL assigned to each target learns the sensing beam in accordance with the description in Section~\ref{secCausalBeam}. Each sensing beam vector undergoes similar steps of perturbation, quantization, and evaluation on the assigned target as that for the communication beam vectors.

\section{CAUSAL DRL FOR BEAM PATTERN DESIGN}
\label{secCausalBeam}
We propose the use of the INVASE framework used in \cite{Sun2022}, to adjust the phase shifters for addressing the beam pattern design problem. The beam design problem aims to either maximize the beamforming gain of communication user or maximizing the beamforming gain for sensing targets, depending on the  of the DRL agent. The RL setup consists of the following building blocks.
\par \noindent $\bullet$ \textbf{State}: We define state $s_t$ as a vector that consists of the phases of all the phase shifters at the $t^{th}$ iteration, $s_t=[\theta_1,\theta_2, ... ,\theta_M ]^T$, where $\theta_1,\theta_2, ... , \theta_M$ are the phase angles of the $M$ BS antennas that constitute the beamforming vector $w_i$ and they lie within the range ($-\pi,\pi$].
\par \noindent $\bullet$ \textbf{Action}: We define action $a_t$ as the element-wise changes to all the phases in $s_t$. The action becomes the next state, i.e., $s_{t+1}=a_t$.
\par \noindent $\bullet$ \textbf{Reward}: We define a reward mechanism for both communication and sensing such that the reward at the $t^{th}$ iteration, denoted as $r_t$, takes values from $\{+1,0,-1\}$. To compute the reward, we compare the current beamforming gain, denoted by $g_t$, with (a) an adaptive threshold $\beta_t$, and (b) the previous beamforming gain $g_{t-1}$. The reward at $t^{th}$ iteration is,
    \begin{align}
    r_t =
        \begin{cases}
            +1, & \text{if } g_t>\beta_t \\
            0, & \text{if } g_t \leq \beta_t  \text{ and } g_t>g_{t-1} \\
            -1, & \text{if } g_t \leq \beta_t  \text{ and } g_t \leq g_{t-1}
        \end{cases}
    \end{align}
The adaptive threshold $\beta_t$ does not rely on any prior knowledge of the channel distribution or the AoA of the sensing target. The threshold has an initial value of zero. Whenever the beamforming gain $g_t$ surpasses the current threshold $\beta_t$, the system updates the threshold with the current gain $g_t$, i.e., $\beta_{t+1}=g_t,\text{ if }r_t=+1$. The sensing beam threshold has a different value than the communication beam threshold. To generalize, there could be other data and domain driven approaches for site-specific reward.

\begin{figure}[htbt!]
\centering
  \includegraphics[width=0.7\linewidth,clip]{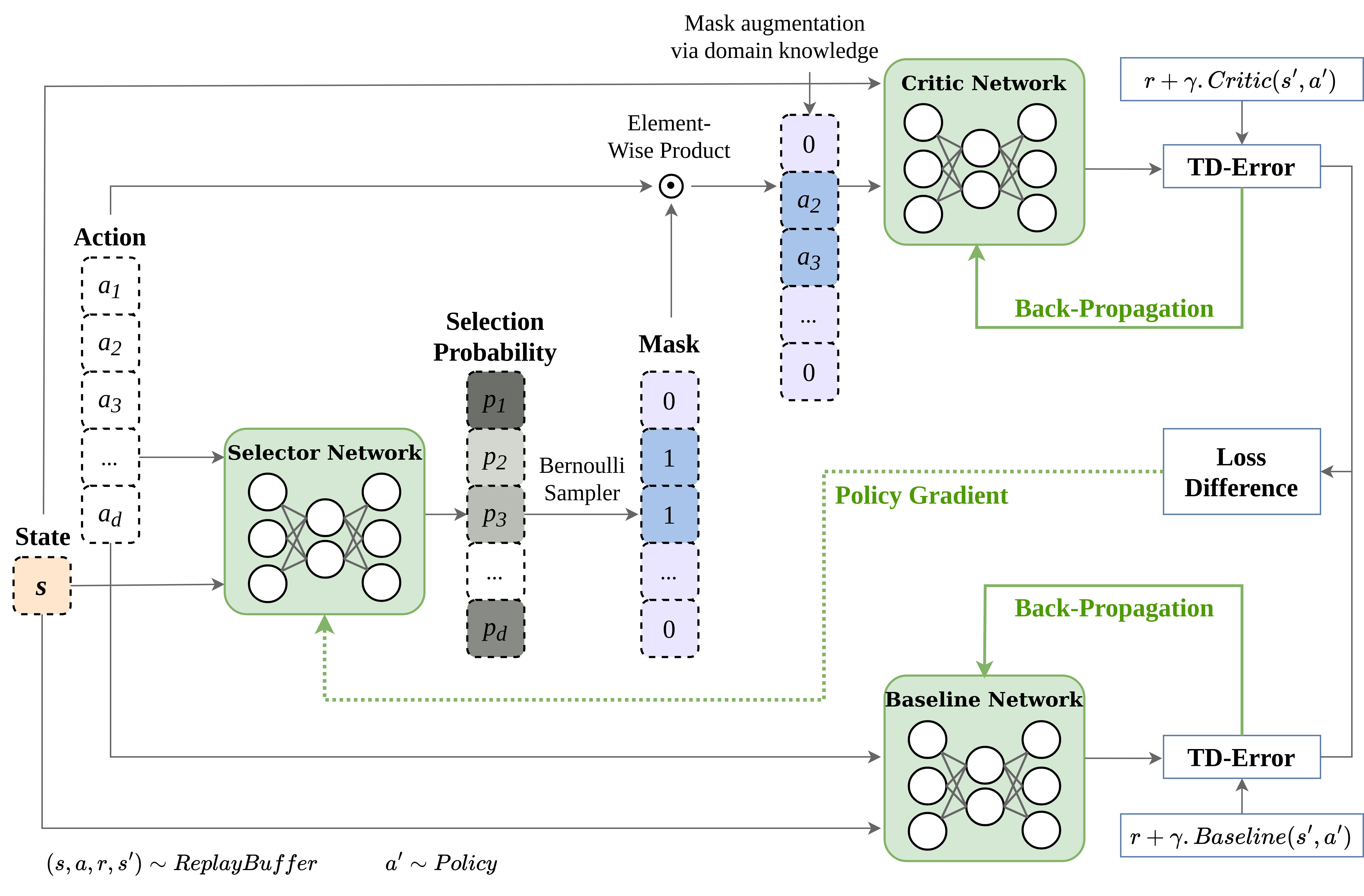}
   \caption{Block diagram of the INVASE architecture present in each DRL agent for beam pattern learning \cite{Sun2022}. The capture of cause-effect relationship can also be augmented by domain knowledge.}
   \label{figInvArch}
\end{figure}

Now, we discuss the vanilla INVASE architecture~\cite{Yoon2018} using temporal difference learning that is used to train our DRL agents. The architecture has been illustrated in Fig.~\ref{figInvArch}. The agent interacts with the environment and stores the state, action, reward, and next state in the replay buffer as a tuple $(s,a,r,s^\prime)$. States and actions sampled from replay buffer are fed into the selector network which selects the selection probabilities of different dimensions of the action. A selection mask is generated according to the selection probability vector and multiplied with the action vector to obtain the most relevant actions. Hence, the selector network intervenes and discovers the causal relationship between the action dimensions and the state. The critic network is trained to minimize the Temporal Difference (TD) error with the states and the selected action dimensions, while the baseline network is trained to minimize the TD error with the states and the primal action space. The difference between the TD errors is used to conduct policy gradient to update the selector network. We use the Iterative Curriculum INVASE (IC-INVASE)~\cite{Sun2022} which incorporates two improvements over vanilla INVASE to make it suitable for action selection when the action dimension in RL is extremely high, for instance, the huge action dimension in mMIMO beamforming applications. The first improvement is based on \emph{curriculum} learning and it is used to tackle the problem of convergence of the INVASE to sub-optimal solutions by pruning all actions when the $\lambda$ parameter in Equation~\ref{eqINVASELoss} is large~\cite{Yoon2018}. The second improvement provides an \emph{iterative} structure of the action selection such that the selection function $G$ (previously defined as $F$ in Section~\ref{secBackgrnd}) conducts hierarchical action selection with less computational expenses.

\subsection{CURRICULUM FOR HIGH DIMENSIONAL ACTION SELECTION}
Curriculum learning~\cite{Bengio2009} mimics human learning by gradually training to make decisions for more difficult tasks and its effectiveness has been demonstrated in several works~\cite{Czarnecki2018,Matiisen2017,Weinshall2018}. Intuitively, it will be easier to select $P$ useful actions out of $L$ primal actions when $P$ is large, the most trivial solution being that all of the $L$ actions are selected, i.e., $x^{(G(x))}=G(x)\odot x=x$. The authors in~\cite{Sun2022} design two curriculum for the purpose of selecting less but important actions, (a) curriculum on the penalty $\lambda$ (Equation~\ref{eqINVASELoss}), and (b) curriculum on the proportion of actions selected. The curriculum on the penalty $\lambda$ is designed in such a way that the $\lambda$ value is increased from 0 to some value, for e.g. $1.0$, to impose larger penalty on the number of actions selected by the selector network $G$. The proportion of the variables selected, denoted as $p_r$, is decreased from a value between $[0,1)$ to $0$ to force the selector to select less but increasingly important actions gradually. For instance, when $p_r=0.5$, the selector network will be penalized for selecting less or more than half of the actions. Thus, the learning objective of the curriculum-INVASE is given as
\begin{align}
    \label{eqLossFunc}
    \nonumber
    \mathcal{L} &= \mathcal{D}_{KL}(p(Y|X=x)||p(Y|X^{G(x)}=x^{G(x)})) \\
    &+ \lambda ||G(x)|_0-d.p_r|
\end{align}

Here, $G:X \to \{0,1\}^d$, $|G(x)|_0$ is the cardinality of the selected actions, $d$ is the primal action dimension, and $x^{G(x)} =G(x) \odot x$ denotes the element-wise product of $x$ and the selection mask $G(x)$.

\subsection{ITERATIVE SELECTION OF ACTIONS}
The selection function $G(x)$ is be applied for multiple times to perform coarse-to-fine action selection. The $i^{th}$ dimension of the action after selection $x_i^{(G(x))}$ is given as
\begin{align}
    x_i^{(G(x))} =
    \begin{cases}
        1, & \text{if } G_i(x) = 1 \\
        0, & \text{if } G_i(x) = 0
    \end{cases}
\end{align}

where $G_i(x)$ is the selection function output for the $i^{th}$ action dimension. Then $x^{(G(x))}$ is fed into the selector network to form the hierarchical action selection over $n$ selection processes given as
\begin{align}
\nonumber
    x^{(1)} &= (G(x) \odot x), \\
    x^{(2)} &= (G(x^{(1)}) \odot x^{(1)}), \\
\nonumber
    \cdots \\
\nonumber
    x^{(n)} &= (G(x^{(n-1)}) \odot x^{(n-1)})
\end{align}

After each selection operation, the most relevant $p_r$ actions are selected, for instance, if $p_r=0.5$, $50\%$, $25\%$, and $12.5\%$ most important actions will be selected after the first three selection operations. IC-INVASE is a combination of these two improvements and this contributes to high learning efficiency in high-dimensional action selection tasks.

\subsection{TEMPORAL DIFFERENCE STATE-WISE ACTION REFINEMENT}
We use the algorithm named TD-SWAR proposed in~\cite{Sun2022} on top of the vanilla Twin Delayed Deep Deterministic Policy Gradient (TD3)~\cite{Scott2018} algorithm (referred to as TD3-INVASE) to minimize the loss function for the IC-INVASE architecture that we have employed for the beam pattern learning by DRL agent (see Figure~\ref{figSysModel}). The pseudocode for the algorithm is presented in Algorithm~\ref{algTDSWAR}, which depicts the steps for updating the parameters of the modules in Fig.~\ref{figInvArch}.

Algorithm~\ref{algTDSWAR} first initializes critic $C_{\phi_1}$, $C_{\phi_2}$, baseline $B_{\psi_1}$, $B_{\psi_2}$, actor $\pi_\textit{v}$, selector $G_\theta$ networks, the corresponding target networks with parameters $\phi^\prime_1$, $\phi^\prime_2$, $\psi^\prime_1$, $\psi^\prime_2$, $\textit{v}^\prime$, $\theta^\prime$, and replay buffer $\mathcal{B}$. At each iteration $t$, the experience tuple is stored in the replay buffer $\mathcal{B}$ as tuple $(s,a,r,s^\prime)$, where $s$ is the agent's state, $a$ is the agent's action, $r$ is the reward obtained from the environment for action $a$, and $s^\prime$ is the predicted next state [Line 3]. For a mini-batch of these tuples sampled from $\mathcal{B}$, the algorithm first perturbs the next action obtained from the target actor $\pi_{v^\prime}$ with noise sampled from Gaussian distribution to give $\Tilde{a}$ [Line 5-6]. The target selector network $G_{\theta^\prime}$ finds relevant actions $\Tilde{a}^{G_{\theta^\prime}(\Tilde{a}|s^\prime)}$ from this perturbed action $\Tilde{a}$ [Line 8]. These relevant action obtained from the selector is used to find the target critic value $y_c$ [Line 8-10], while the target baseline value $y_b$ is found using the entire perturbed action $\Tilde{a}$. The critic and baseline networks are updated with the Mean-Square Error (MSE) loss between the critic and its target value, and baseline and its target value [Line 13-15]. Here, the critic's output is obtained using the selector's output $a^{G(a|s)}$ on the current action $a$. The difference between the MSE losses of critic and baseline gives the loss difference (see Figure~\ref{figInvArch}) and this loss difference is used conduct the policy gradient to update the selector network [Line 17]. The actor is updated using policy gradient method with different learning rate [Line 20], after which, the target networks are updated [Line 22-24].

\begin{algorithm}[!htbt]
\footnotesize{
\caption{TD-SWAR}\label{algTDSWAR}
\begin{algorithmic}[1]
\STATE Initialize critic networks $C_{\phi_1}$, $C_{\phi_2}$, baseline networks $B_{\psi_1}$, $B_{\psi_2}$ and actor network $\pi_\textit{v}$, selector network $G_\theta$, target networks $\phi^\prime_1 \gets \phi_1$, $\phi^\prime_2 \gets \phi_2$, $\psi^\prime_1 \gets \psi_1$, $\psi^\prime_2 \gets \psi_2$, $\textit{v}^\prime \gets \textit{v}$, $\theta^\prime \gets \theta$, replay buffer $\mathcal{B}$
\STATE \textbf{for} $t=1,H$ \textbf{do}
\STATE \hspace{0.5cm} Interact with the environment and store transition $(s,a,r,s^\prime)$ in $\mathcal{B}$
\STATE \hspace{0.5cm} Sample mini-batch of transitions from $\mathcal{B}$
\STATE \hspace{0.5cm} Sample $\epsilon$ from a clipped Gaussian distribution
\STATE \hspace{0.5cm} Calculate perturbed next action $\Tilde{a} \gets \pi_{\textit{v}^\prime}(s^\prime) + \epsilon$
\STATE \hspace{0.5cm} Select actions with the target selector network
\STATE \hspace{0.7cm} $\Tilde{a}^{G_{\theta^\prime}(\Tilde{a}|s^\prime)} \gets G_{\theta^\prime}(\Tilde{a}|s^\prime) \odot \Tilde{a}$
\STATE \hspace{0.5cm} Calculate the target critic and baseline values, $y_c$ and $y_b$
\STATE \hspace{0.7cm} $y_c \gets r + \gamma \min_{i=1,2} C_{\phi^\prime_i}(s^\prime, \Tilde{a}^{G_{\theta^\prime}(\Tilde{a}|s^\prime)})$
\STATE \hspace{0.7cm} $y_b \gets r + \gamma \min_{i=1,2} B_{\psi^\prime_i}(s^\prime, \Tilde{a})$
\STATE \hspace{0.5cm} Update critics and baselines
\STATE \hspace{0.7cm} $a^{G_\theta(a|s)} \gets G_{\theta}(a|s) \odot a$
\STATE \hspace{0.7cm} $\phi_i \gets \arg\min_{\phi_i} \textbf{MSE}(y_c, C_{\phi_i}(s,a^{G_\theta(a|s)}))$
\STATE \hspace{0.7cm} $\psi_i \gets \arg\min_{\psi_i} \textbf{MSE}(y_b, B_{\psi_i}(s,a))$
\STATE \hspace{0.5cm} Update selector network by policy gradient (learning rate $\eta_1$) 
\STATE \hspace{0.7cm} $\theta \gets \theta - \eta_1(l_c - l_b)\nabla_\theta \log G_\theta(a|s)$, $l_c$, $l_b$ are MSE losses from 
\STATE \hspace{0.7cm} previous step
\STATE \hspace{0.5cm} Update actor network by policy gradient (learning rate $\eta_2$) 
\STATE \hspace{0.7cm} $\theta \gets \theta - \eta_2\nabla_a C_{\phi_1}(s,a)|_{a=\pi_\textit{v}(s)} \nabla_\textit{v} \pi_\textit{v}(s)$
\STATE \hspace{0.5cm} Update target network with $\tau \in (0,1)$
\STATE \hspace{0.7cm} $\phi^\prime_i \gets \tau\phi_i + (1-\tau)\phi^\prime_i$
\STATE \hspace{0.7cm} $\psi^\prime_i \gets \tau\psi_i + (1-\tau)\psi^\prime_i$
\STATE \hspace{0.7cm} $\textit{v}^\prime_i \gets \tau\textit{v}_i + (1-\tau)\textit{v}^\prime_i$
\STATE \textbf{end}
\end{algorithmic}
}
\end{algorithm}

\section{EXPERIMENTAL SETUP}
\label{secExpSetup}
In this section, we will discuss about the channel dataset generation from the training scenario used in our experiments, followed by the training setup.

\subsection{DATASET GENERATION FOR JCAS}
\begin{figure}[htbt!]
  \centering \includegraphics[width=0.7\linewidth,clip]{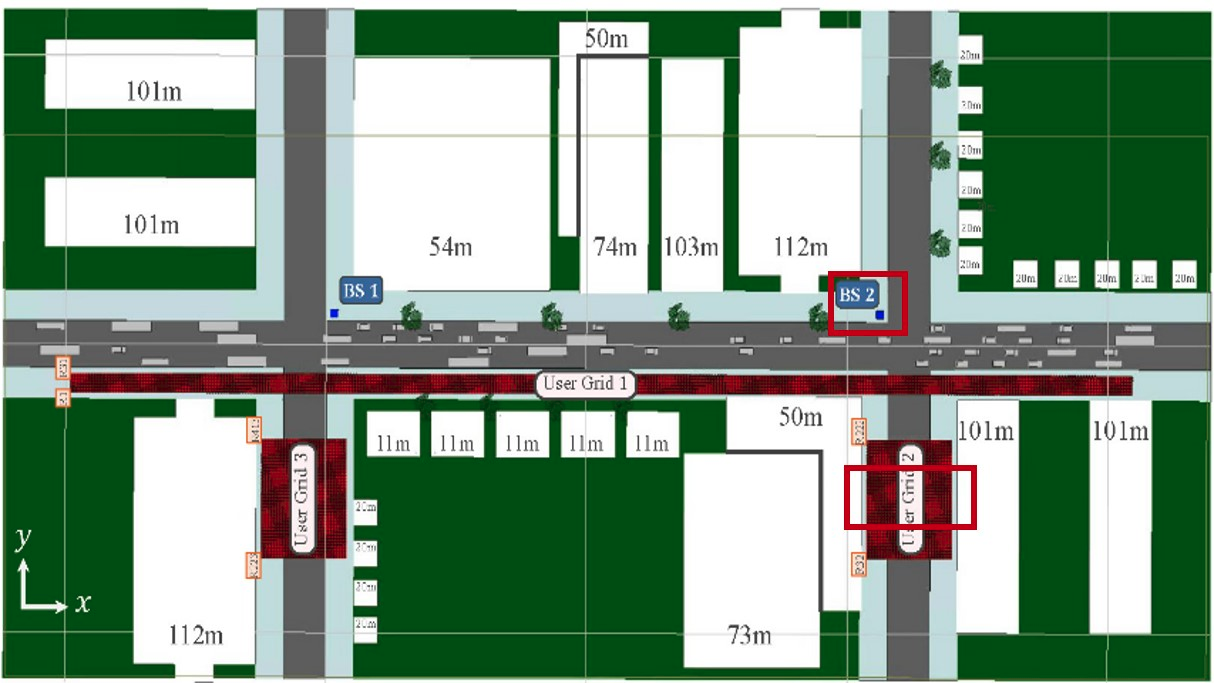}
   \caption{Top view of the dynamic scenario for training the deep reinforcement learning agents~\cite{DeepMIMOwebsite}. The active base station and user grid for generating the training data are indicated with \emph{red boxes}.}
   \label{figTrainScenario}
\end{figure}

To conduct the performance evaluation of the proposed TD3-INVASE for beamforming in a JCAS setup, we consider a dynamic scenario where the base station serves the communication among static users while sensing the vehicles nearest to the communicating users. In our experiments, we demonstrate JCAS at the base station by learning beams for communicating users and beams for sensing vehicles nearest to the communicating users. Therefore, we require the channel dataset (vectors) for training communication beams, and the locations of vehicles, users and base station (coordinates) to compute the AoA for training sensing beams. The channel, vehicle and base station location datasets are sampled from the dynamic scenario, as shown in Figure~\ref{figTrainScenario}, at 100 ms and 1000 such sampled snapshots are available as a part of the MIMO channel dataset obtained from benchmarking channel generation tool DeepMIMO~\cite{Alkhateeb2019}.  

The dynamic scenario consists of an urban road network where there are communicating users and vehicles moving on the road in between the users and the base station. There are three grids of communicating users and two base stations, out of which the selected user grid and the active base station for our experiment have been highlighted in \emph{red boxes} in Figure~\ref{figTrainScenario}. Using DeepMIMO scripts, we generate two channel datasets $\mathcal{H}_{AR^\prime}$ and $\mathcal{H}_{AR}$ for the selected user grid and active base station, where $\mathcal{H}_{AR^\prime}$ and $\mathcal{H}_{AR}$ indicates channel dataset \emph{without antenna rotation} and \emph{with antenna rotation} respectively. Table~\ref{tableChGenParam} shows the channel data generation hyper-parameters. The antenna orientation angles ($\theta_x,\theta_y,\theta_z$) represent the angle of rotation along x, y, and z axes. In order to generate $\mathcal{H}_{AR^\prime}$, we set the ($\theta_x,\theta_y,\theta_z$) angles to (0,0,0) for both the BS and UEs, while for generating $\mathcal{H}_{AR}$, we set them to (0,30,0) for BS and a random value ranging between [0,30], [30,60], and [60,90] for $\theta_x$, $\theta_y$, and $\theta_z$ in case of UEs. From the location dataset, we find the vehicle closest to the selected user grid at each sampled scenario by using $j_{min} = \underset{j}{argmin} \text{ } dist(user(i)-veh(j))$, where $j_{min}$ is the index of the vehicle closest to the selected user distribution. The channel datasets $\mathcal{H}_{AR^\prime}$ and $\mathcal{H}_{AR}$ for each sampled scenario are given as input to the beam codebook learning framework for JCAS to learn communication beams as described in Section~\ref{secBeamCodebook}, while the locations of the vehicle index $j_{min}$ at each sampled scenario is given as sensing target input (dataset) to learn sensing beams.

\begin{table}[htbt!]
    \centering
        \begin{tabular}{p{4.2cm} p{3cm}}
        \hline
        \textbf{Parameter} & \textbf{Value} \\
        \hline
        \hline
        Name of scenario & O2\_dyn\_3p5 \\
        Active BS & 2 \\
        Active users & 100 to 150 \\
        \hline
        Number of antennas (x,y,z) & (1,32,1) \\
        BS antenna orientation ($\theta_x,\theta_y,\theta_z$) & \{(0,0,0), (0,30,0)\} \\
        UE antenna orientation ($\theta_x,\theta_y,\theta_z$) & \{(0,0,0), ([0,30],[30,60],[60,90])\} \\
        \hline
        System BW & 0.5 GHz \\
        Antenna spacing & 0.5 m \\
        Number of OFDM sub-carriers & 1 \\
        \hline
        OFDM sampling factor & 1 \\
        OFDM limit & 1 \\
        Number of multipaths & 5 \\
        \hline
        \end{tabular}
    \vspace{0.4 cm}
    \caption{Channel generation hyper-parameters for training}
    \label{tableChGenParam}
\end{table}

\subsection{TRAINING MODEL}
In all experiments performed in the subsequent section, we use the standard neural network structure of actor and critic. Baseline Deep Deterministic Policy Gradient (DDPG)~\cite{lillicrap2015} algorithm is used to train this neural network structure. The input to the actor network is the state, i.e. phases of the antenna phase shifters, and hence has a dimension of $M$. The actor has two hidden layers of $16M$ neurons which are followed by Rectified Linear Unit (ReLU) activations. The actor's output layer is activated using hyperbolic tangent (tanh) scaled by $\pi$ to give the predicted action of dimension $M$. The critic network takes as input the concatenation of state and action with dimension $2M$ and has two hidden layers with $32M$ neurons followed by ReLU activations. The critic's output is a 1-dimensional predicted Q value of the input state-action pair. Baseline TD3 used in our experiments has two critics, each having two hidden layers of $32M$ neurons followed by ReLU activations. For TD3-INVASE, the two critics take as input the $2M$ dimensional state-selected actions pair, while the baseline networks takes as input the $2M$ dimensional state-primal actions pair. Both the critic and baseline networks have two hidden layers of $32M$ neurons followed by ReLU activations and each of them give 1-dimensional Q value output. The selector network have two hidden layers of $100$ neurons and they are followed by sigmoid activation. 

In our experiments, we have $M$ antennas on the base station, and hence state and action dimensions are $M$. We add $M_{Red}$ fictitious antennas to these actual antennas, which is analogous to the fact that we are injecting $M_{Red}$ redundant actions into the action space. These redundant action dimensions will not affect the state transitions or the reward calculation, but the learning agent needs to identify the redundant actions among the $M+M_{Red}$ actions by finding the state-action causal relationships and pruning them during training in order to perform the learning process efficiently.

\section{RESULTS}
\label{secResults}


\begin{table}[htbt!]
\centering
       \begin{tabular}{p{3.2cm} p{0.8cm}}
        \hline
        \textbf{Parameter} & \textbf{Value} \\
        \hline
        \hline
        Mini-batch size & 1024 \\
        Learning rate & $10^{-3}$ \\
        \hline
        Number of epochs & 200000 \\
        Data buffering epochs & 50000 \\
        Epochs per episode & 200 \\
        \hline
        State dimension ($M$) & 32 \\
        Action dimension ($M$) & 32 \\
        Redundant actions ($M_{Red}$) & 100 \\
        \hline
        \end{tabular}
    \caption{Model training hyper-parameters}
    \label{tableHypParam}
\end{table}

We evaluate the performance of our proposed learning model for JCAS by using the two metrics, (a) average Episodic Beamforming Gain, and (b) True Positive Rate (TPR). The average Episodic Beamforming Gain is computed as the average of the beamforming gain $g_t$ over each episode. Let us consider that the number of iterations/epochs in each episode is $E$. Then, the average Episodic Beamforming Gain for communication beam is $g_t=\frac{1}{E}\sum_{t=1}^{E}|w_t^Hh_u|^2$, where $w_t$ is the beamforming vector at $t^{th}$ iteration and $h_u$ is the channel vector. For sensing beam, the same metric is computed as $g_t=\frac{1}{E}\sum_{t=1}^{E}|w_t^Hb_r|^2$, where $b_r$ is the antenna steering vector (Equation~\ref{eqAntArray}). In order to demonstrate the efficacy of the selection process for relevant actions by TD3-INVASE, we define the TPR metric as the percentage ratio of the number of selected actions $|G|$ to the number of actual relevant actions $M$. Here, we assume that we have knowledge of the actual relevant actions for the mMIMO system beforehand. The closeness of $|G|$ to $M$ quantifies how well the TD3-INVASE model is able to prune the irrelevant action dimensions, for instance, a TPR of $50\%$ denotes that the TD3-INVASE is able to select only half of the relevant action dimensions.

\subsection{TRAINING RESULTS}
\begin{figure}[htbt!]
    \centering    \includegraphics[width=0.5\columnwidth,clip]{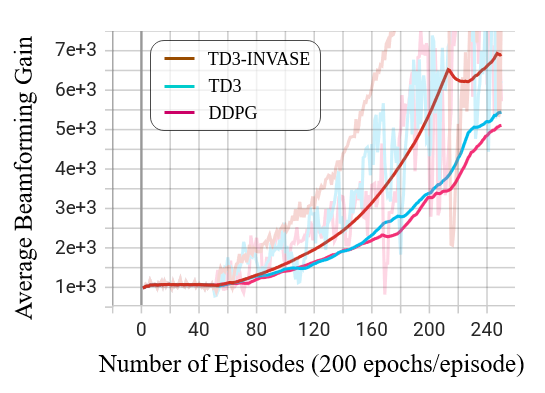}
    \caption{Average episodic beamforming gain for communication beam in scene 0.}
    \label{figTD3INVvsBase}
\end{figure}


We provide quant-itative comparison between TD3-INVASE, and baselines vanilla TD3 and DDPG policies for learning communication and sensing beams by training the DRL agents on the channel and sensing datasets. The hyper-parameters used for training can be found in Table~\ref{tableHypParam}. In mMIMO systems, the number of antennas on the transmitter/receiver units are higher than 16~\cite{Feng2019}. Therefore, we use antenna array of size $M+M_{Red}=132$ for the mMIMO system in our experiments. 

The first result shows the performance in terms of the average beamforming gain versus the number of episodes comparison, as depicted in Fig.~\ref{figTD3INVvsBase}, where each episode consists of 200 epochs. Fig.~\ref{figTD3INVvsBase} clearly depicts that the agent learning with TD3-INVASE outperforms both the vanilla TD3 and DDPG baselines. The higher beamforming gain for TD3-INVASE indicates that it is reaching the maximum reward in the minimum number of training epochs per episode, and it is reaching this maximum reward in fewer episodes than the baselines. This indicates that TD3-INVASE is sample efficient and learns faster than TD3 and DDPG to reach maximum reward in each episode.


\begin{figure}[htbt!]
  \centering \includegraphics[width=0.7\linewidth,clip]{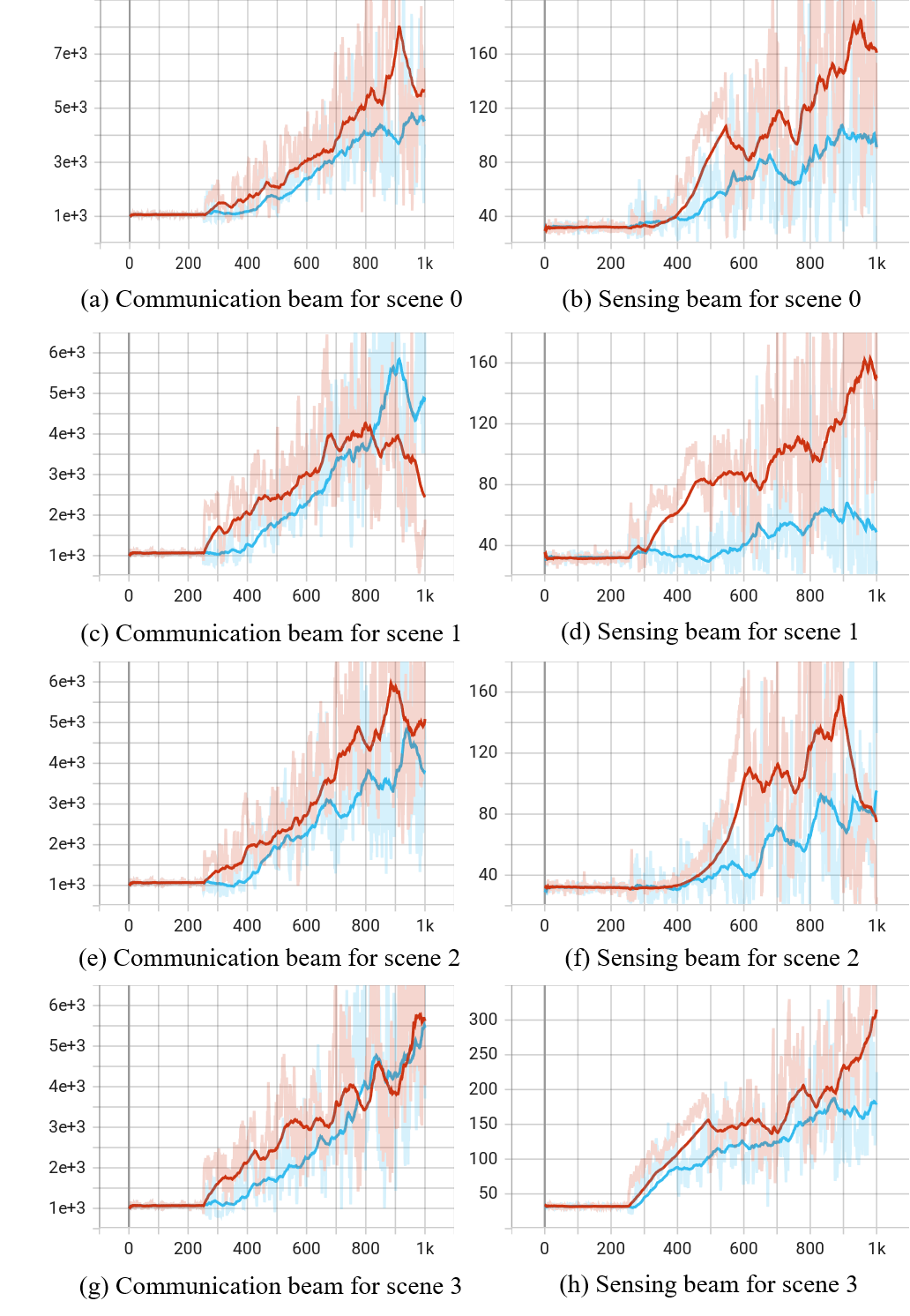}
   \caption{Average episodic beamforming gain for JCAS \emph{without antenna rotation} in scene 0 ((a)-(b)), scene 1 ((c)-(d)), scene 2 ((e)-(f)), and scene 3 ((g)-(h)). The \emph{brown} and \emph{blue} plots indicate the average beamforming gain for TD3-INVASE and TD3 respectively.}
   \label{figTrainResults}
\end{figure}

\begin{figure*}[htbt!]
  \centering \includegraphics[width=0.8\linewidth,clip]{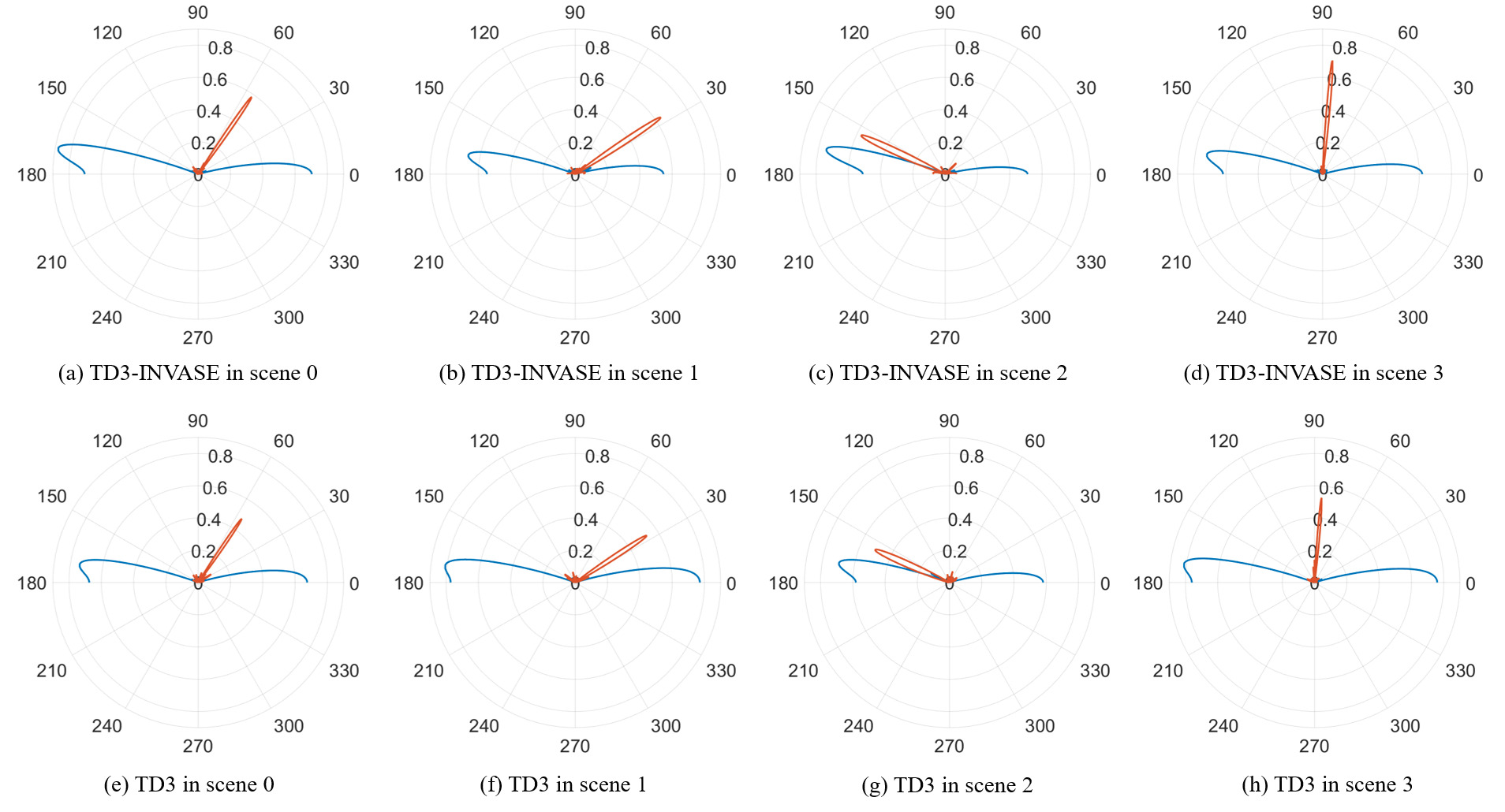}
   \caption{Trained beam pattern for JCAS \emph{without antenna rotation} in scene 0 ((a)-(e)), scene 1 ((b)-(f)), scene 2 ((c)-(g)), and scene 3 ((d)-(h)). The \emph{blue} and \emph{brown} beam patterns indicate the communication and sensing beams respectively.}
   \label{figTrainBeam}
\end{figure*}

\subsubsection{WITHOUT ANTENNA ROTATION} 
The set of results depicted in Fig.~\ref{figTrainResults} shows the performance (Average Beamforming Gain versus Number of Episodes) of TD3-INVASE and TD3 for both communication and sensing in 4 consecutive scenes (referred to as scene 0, 1, 2, and 3), where antennas on BS and UEs have no rotation. In all the scenes and for both communication and sensing, the agent learning with TD3-INVASE outperforms vanilla TD3 baseline by achieving higher gain in fewer episodes. The DRL agents associated with each of the communication and sensing beams selects the relevant actions by using the TD3-INVASE architecture, thus maximizing the reward per episode which translates to higher beamforming gain in fewer episodes indicating sample efficiency. Therefore in all the scenes, the TD3-INVASE architecture is able to achieve high sample efficiency.

Fig.~\ref{figTrainBeam} depicts the beam pattern for JCAS after the learning process is complete. In each plot, the normalized gain is indicated by the concentric circles, while the values along the circumference of the outermost circle indicate the angle of the formed beam. The \emph{blue beam} is the communication beam, while the \emph{brown beam} is the sensing beam. It shows that the proposed TD3-INVASE model forms communication and sensing beams with higher gain and better directivity than the vanilla TD3 baseline. The high beamforming gain achieved by the TD3-INVASE architecture gets reflected in the beam pattern plot as beams reaching higher normalized gain, for example, in Fig.~\ref{figTrainBeam}(a) and (b) (plots for scene 1), TD3-INVASE achieves a normalized gain of more than 0.8 for the communication beam, while the vanilla TD3 achieves normalized gain less than 0.8. Similarly, in the same figures, the sensing beam reaches 0.6 normalized gain using TD3-INVASE, which is higher than that for its vanilla TD3 implementation.

\begin{figure}[htbt!]
  \centering \includegraphics[width=0.7\linewidth,clip]{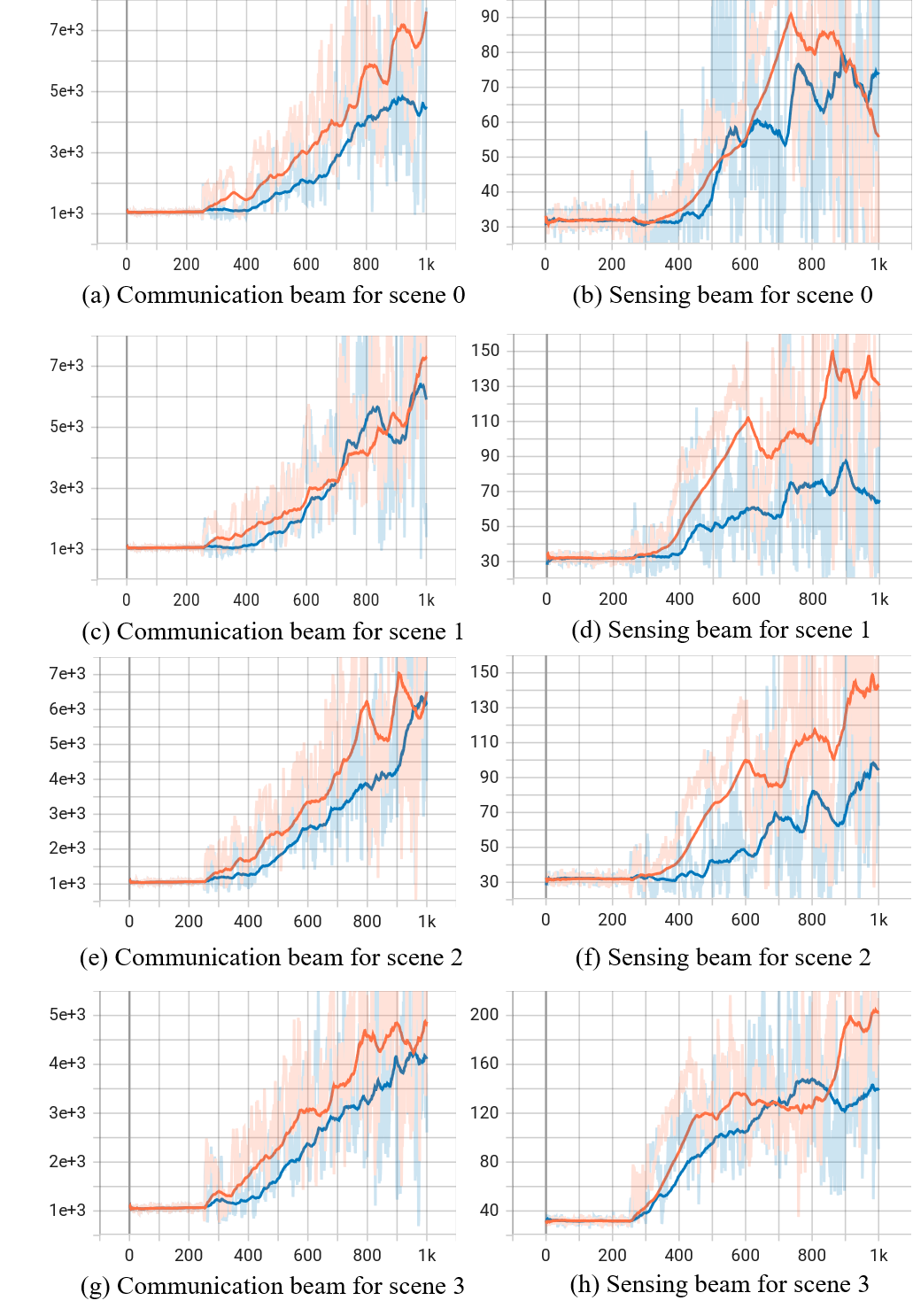}
   \caption{Average episodic beamforming gain for JCAS \emph{with antenna rotation} in scene 0 ((a)-(b)), scene 1 ((c)-(d)), scene 2 ((e)-(f)), and scene 3 ((g)-(h)). The \emph{orange} and \emph{blue} plots indicate the average beamforming gain for TD3-INVASE and TD3 respectively.}
   \label{figTrainResultsAntOrient}
\end{figure}

\begin{figure*}[htbt!]
  \centering \includegraphics[width=0.8\linewidth,clip]{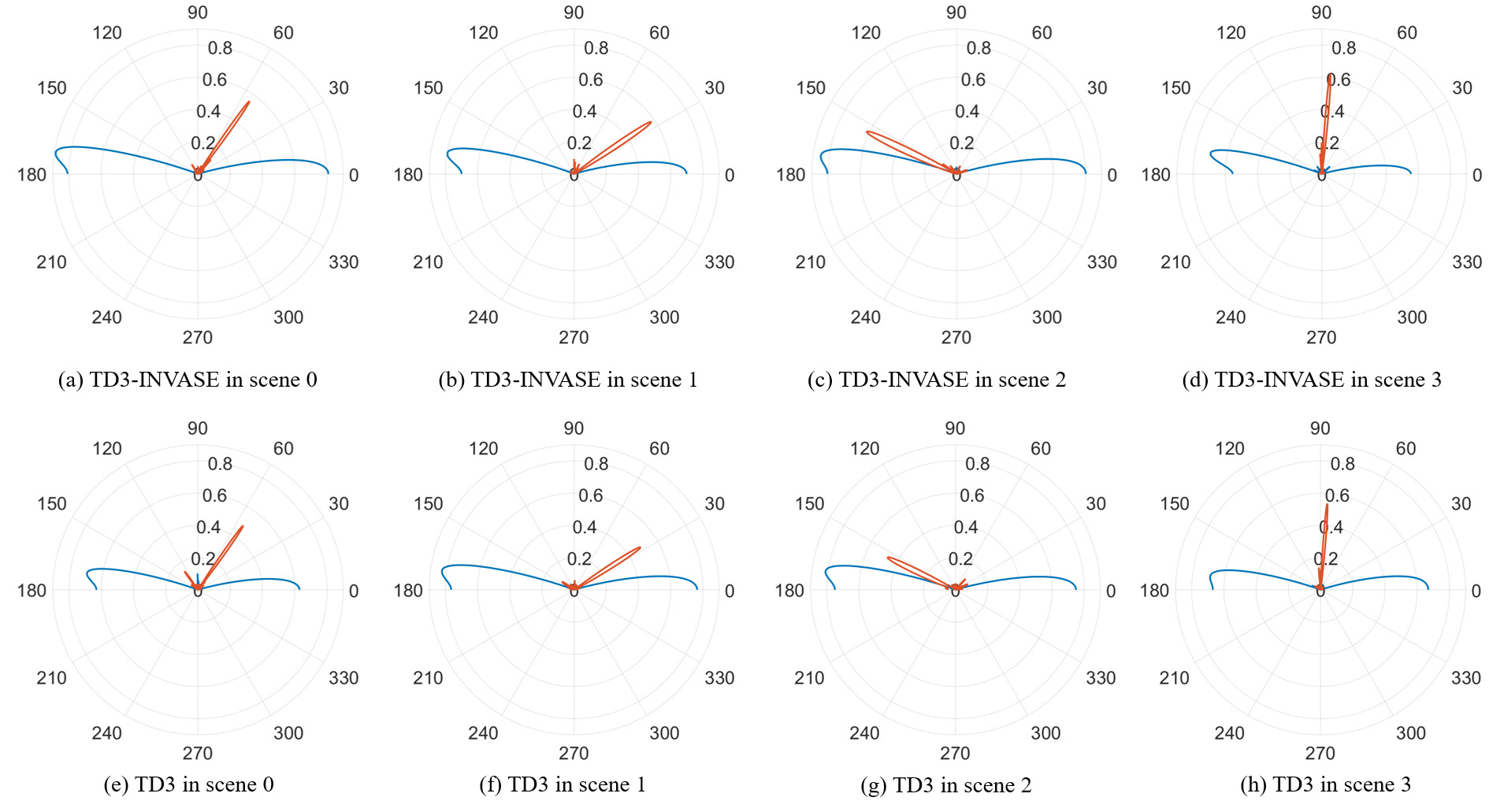}
   \caption{Trained beam pattern for JCAS \emph{with antenna rotation} in scene 0 ((a)-(b)), scene 1 ((c)-(d)), scene 2 ((e)-(f)), and scene 3 ((g)-(h)). The \emph{brown} and \emph{blue} beam patterns indicate the average beamforming gain for TD3-INVASE and TD3 respectively.}
   \label{figTrainBeamAntOrient}
\end{figure*}

\subsubsection{WITH ANTENNA ROTATION}
Here, we analyse the training results illustrated in Fig.~\ref{figTrainResultsAntOrient}. It shows the performance of TD3-INVASE and TD3 for both communication and sensing in the same 4 consecutive scenes, but with the antennas on BS and UEs having rotation (refer to Table~\ref{tableHypParam}). It can be observed that in all the scenes for both communication and sensing, the agent with TD3-INVASE achieves higher average gain in fewer episodes when compared to vanilla TD3. This proves that even when there are different antenna configurations of the BS and UEs, the TD3-INVASE model selects the relevant actions and achieve better sample efficiency than TD3.

The trained beam pattern plots depicted in Fig~\ref{figTrainBeamAntOrient} show that TD3-INVASE model forms communication and sensing beams with higher gain and directivity than the TD3 baseline even when the BS and UEs have different antenna rotation. The higher beamforming gain leads to higher normalized gain in the beam pattern plots for both communication and sensing. 

Table~\ref{tableTPR} shows the quantitative performance of TD3-INVASE in terms of TPR metric. TD3-INVASE achieves high TPR of $100\%$ for communication beam learning in most of the experiments considered in our study. The TPR for sensing beam has a moderate performance of $70\%$ on an average. This indicates that the model is able to successfully select most of the relevant actions from among the primal actions. Therefore, we can infer from the training results that the proposed TD3-INVASE model for JCAS can achieve higher sample efficiency for different antenna configurations and different user environments, while successfully selecting most of the relevant actions from the primal action space.

\begin{table}[htbt!]
    \centering
        \begin{tabular}{p{0.5cm} p{1.2cm} p{1.1cm} p{1.2cm} p{1.1cm}}
        \hline
        & \multicolumn{2}{l}{\textbf{Without Antenna Rotation}} & \multicolumn{2}{l}{\textbf{With Antenna Rotation}} \\ 
        \hline
        \textbf{Scene} & \textbf{Comm-unication} & \textbf{Sensing} & \textbf{Comm-unication} & \textbf{Sensing} \\
        \hline
        \hline
        \textbf{0} & 100 & 53.2 & 100 & 65.6 \\
        \textbf{1} & 90.6 & 71.8 & 100 & 68.7 \\
        \textbf{2} & 93.7 & 75 & 93.7 & 78.1 \\
        \textbf{3} & 100 & 71.8 & 100 & 81.2 \\
        \hline
        \end{tabular}
    \vspace{0.4 cm}
    \caption{TPR for communication and sensing over 1000 episodes of TD3-INVASE for BS antenna orientations (without and with antenna rotation) in different scenes.}
    \label{tableTPR}
\end{table}



\subsection{TESTING RESULTS}

Now, we conduct testing of the trained model for different channel datasets obtained by re-orienting the BS antennas to different antenna rotations $(\theta_x,\theta_y,\theta_z)$ for the scenario depicted in Fig.~\ref{figTrainScenario}. Our trained has no knowledge about these datasets. Table~\ref{tableAntTilt} depicts the average beamforming gain attained by the trained TD3 and TD3-INVASE models for different antenna tilts. It can be observed that for the sensing beam, TD3-INVASE models gives more gain than TD3 over all the antenna tilts. Here, the the gain for sensing beam is unaffected by the antenna tilt as the gain and direction of the sensing beam depends only on the AoA of the sensing beam reflected from the target and the spacing between the mMIMO antenna elements of the base station, which are constant (Refer to Sec.~\ref{secProbForm}). The communication beam undergoes degradation in the gain value as the antenna tilt over each of the $x$ and $z$ axes are increased. For smaller antenna tilts, i.e., (15,0,0), (30,0,0), (0,0,15), and (0,0,30), the TD3-INVASE model achieves significantly higher gains compared to the TD3 model. However, for larger tilts, i.e, (60,0,0), (90,0,0), (0,0,60), and (0,0,90), the gain achieved by TD3-INVASE degrades and TD3 achieves higher gain compared to TD3-INVASE. 

\begin{table}[htbt!]
    \centering
        \begin{tabular}{p{1cm} p{1cm} p{1cm} p{1cm} p{1cm}}
        \hline
        & \multicolumn{2}{l}{\textbf{Communication Beam}} & \multicolumn{2}{l}{\textbf{Sensing Beam}} \\ 
        \hline
        \textbf{Antenna Tilt} & \textbf{TD3} & \textbf{TD3-INVASE} & \textbf{TD3} & \textbf{TD3-INVASE} \\
        \hline
        \hline
        \textbf{(15,0,0)} & 1969.3 & \textbf{8328.1} & 21.7 & \textbf{25.3}\\
        \textbf{(30,0,0)} & 642.2 & \textbf{1214.7} & 21.7 & \textbf{25.3}\\
        \textbf{(60,0,0)} & \textbf{327.1} & 162.5 & 21.7 & \textbf{25.3}\\
        \textbf{(90,0,0)} & \textbf{315.8} & 247.4 & 21.7 & \textbf{25.3}\\
        \hline
        \textbf{(0,15,0)} & 3214.5 & \textbf{12788.4} & 21.7 & \textbf{25.3}\\
        \textbf{(0,30,0)} & 3214.5 & \textbf{12788.4} & 21.7 & \textbf{25.3}\\
        \textbf{(0,60,0)} & 3214.5 & \textbf{12788.4} & 21.7 & \textbf{25.3}\\
        \textbf{(0,90,0)} & 3214.5 & \textbf{12788.4} & 21.7 & \textbf{25.3}\\
        \hline
        \textbf{(0,0,15)} & 3686.7 & \textbf{15122.5} & 21.7 & \textbf{25.3}\\
        \textbf{(0,0,30)} & 2369.7 & \textbf{8974.5} & 21.7 & \textbf{25.3}\\
        \textbf{(0,0,60)} & \textbf{675.8} & 641.1 & 21.7 & \textbf{25.3}\\
        \textbf{(0,0,90)} & \textbf{1232.6} & 338.8 & 21.7 & \textbf{25.3}\\
        \hline
        \end{tabular}
    \vspace{0.4 cm}
    \caption{Average beamforming gain for JCAS over 250 episodes of baseline and TD3-INVASE for different BS antenna orientations in scene 0. The best gain for each beam is highlighted in bold.}
    \label{tableAntTilt}
\end{table}

TD3-INVASE discovers causal relationships by choosing less and relevant actions for the task in hand. However, the causal relationships is not stationary across environments such as in this case of increasing antenna tilts. For higher degrees of antenna tilt, more action dimensions which are not being considered by the trained TD3-INVASE model have more relevance to achieve higher gain. Therefore, the trained TD3 model, which considers all the actions for the gain computation, incorporates higher number of relevant actions required for the highly degraded antenna tilt environments. This makes TD3 model perform better in higher antenna tilt environments. Therefore, as our future work, we will explore the generalization of TD3-INVASE over all scenarios that our framework has not interacted with before.

\section{CONCLUSION}
\label{secConc}
In this paper, we tackle the problem of action exploration with respect to the large action space in mMIMO JCAS setup. We employ intervention to discover the causal relationship between action and reward, and prune the action space so that the most relevant actions are used to achieve the maximum reward. We propose a RL-based framework for JCAS which prunes the most relevant actions by using an action selection network in its learning architecture, referred to as TD3-INVASE. The training results show that our proposed method is sample efficient over different scenarios. In case there is a need for retraining to consider certain adjustments, the proposed model is amenable to online learning. As future work, we plan to generalize our proposed RL framework for environments that it has not interacted with from beforehand. We plan to employ causal or counterfactual reasoning to show that our RL framework generalizes over unknown environments.


\bibliographystyle{unsrt}
\bibliography{references}

\end{document}